\documentclass[aps,prl,twocolumn
]{revtex4}
\usepackage[dvips]{graphicx}

\begin{document}
\title{Gate Capacitance Coupling of Singled-walled Carbon Nanotube
Thin-film Transistors}
\author{Qing Cao, Minggang Xia\thanks{Current address: Department of Physics, Xian Jiaotong
University, Xian, China}, Coskun Kocabas, Moonsub Shim, John A.
Rogers}
 \affiliation{
Department of Materials Science and Engineering, Department of
Physics, Department of Mechanical Science and Engineering,
Department of Electrical and Computer Engineering, Department of
Chemistry, Beckman Institute for Advanced Science and Technology
and Frederick Seitz Materials Research Laboratory at the
University of Illinois at Urbana--Champaign, 405 N.Mathews Ave.,
Urbana, IL 61801, USA }

\author{ Slava V. Rotkin}
\affiliation{Department of Physics and Center for Advanced
Materials and Nanotechnology, Lehigh University, 16 Memorial
Dr.E., Bethlehem, PA 18015, USA}

\begin{abstract}
The  electrostatic  coupling between  singled-walled  carbon
nanotube  (SWNT) networks/arrays and planar gate  electrodes in
thin-film  transistors (TFTs) is analyzed  both  in  the quantum
limit with an analytical model and in the classical limit    with
finite-element   modeling.    The   computed capacitance  depends
on  both the  thickness  of  the  gate dielectric  and the average
spacing between the tubes,  with some  dependence  on  the
distribution  of  these  spacings. Experiments  on  transistors
that use sub-monolayer,  random networks   of   SWNTs  verify
certain  aspects   of   these calculations.  The results are
important for the development of  networks or arrays of nanotubes
as active layers in TFTs and other electronic devices.
\end{abstract}

\maketitle

       Single-walled  carbon nanotube (SWNT) networks/arrays
show great promise as active layers in thin-film transistors
(TFTs).\cite{1}   Considerable progress has been made in  the last
couple  of  years  to  improve  their  performance,  and  to
integrate  them with various substrates, including  flexible
plastics.\cite{2,3}  Nevertheless, the understanding the physics
of the  electrostatic coupling between SWNT networks/arrays and
the  planar  gate  electrode, which is  critical  to  device
operation   and  can  be  much  different   than   that   in
conventional  TFTs, is not well established.   The  simplest
procedure, which has been used in many reports of SWNT TFTs, is to
treat  the  coupling as that  of  a  parallel  plate capacitor,
for which the gate capacitance, $C_i$, is  given  by
$\epsilon/4\pi d$, where $\epsilon$ and $d$ are the dielectric
constant  and the thickness  of gate dielectric, respectively.
This procedure enables  a  useful approximate evaluation  of
device  level performance,\cite{4,5,6}   but   it   is
quantitatively incorrect especially when the average spacing
between SWNTs  is  large compared  to  the  thickness of the
dielectric.\cite{7}   Since  $C_i$ critically  determines  many
aspects  of  device  operation, accurate  knowledge of this
parameter is important both  for device optimizing device designs
and for understanding basic transport mechanisms in the SWNT
networks/arrays.

   In  this  letter, we use a model system,  illustrated  in
Fig.  1,  consisting of a parallel array  of  evenly  spaced SWNTs
fully embedded in a gate dielectric with a planar gate electrode
to examine capacitive coupling in SWNT TFTs.   The influence of
nonuniform intertube spacings is also evaluated to  show the
applicability of those results to real devices. Results  obtained
in the single subband quantum  limit  and those obtained in the
classical limit agree qualitatively in the  range of dielectric
thicknesses and tube densities  (as measured in number of tubes
per unit length) explored  here. The  models are used to
 provide
insights  into  factors  that limit  the  effective  mobilities
($\mu$) achievable  in  these devices.

   We  begin  by calculating $C_i$ of the model system  in  the
quantum limit, where the charge density of the SWNT  in  the
ground state has full circular symmetry.\cite{8}  In this case,
the charge  distributes itself uniformly around  the  tube,  and
each tube, when tuned to the metallic region, can be treated as  a
perfectly conducting wire with radius $R$  and  uniform linear
charge  density  $\rho$.   The  electrostatic  potential induced
by such a perfect conducting wire is: $\phi(r)=2\rho\log(R/r)$
where $r$ is the  distance  away  from  the  center  of  the wire.
The potential  generated by an array of such  wires  is  also  a
linear  function of r and in general, the induced  potential at
the  i-th  tube  depends on $\rho$ of  every  tube  in  array
according to
\begin{equation}
\phi^{ind}\left([\rho_j]\right)= \sum_j C_{ij}^{-1} \rho_j
\label{eq1}
\end{equation}
where coefficients of inductive coupling between i-th and j-th
tube $C_{ij}^{-1}$ are geometry dependent.\cite{9}

   For  SWNTs  in the metallic regime, $\rho$ is proportional  to
the shift of the Fermi level which is itself proportional to an
average acting potential at the nanotube according to:
\begin{equation}
\rho=-C_Q \phi^{act}=-C_Q \left(\phi^{xt}+ \phi^{ind}\right)
\label{eq2}
\end{equation}
where   the  proportionality  coefficient  is  the   quantum
capacitance $C_Q$.\cite{8,10,11}  This equation is written to
separate contributions from the external potential, $\phi^{xt}$,
as applied by distant electrodes and as generated by any other
external source associated with charge traps, interface states,
etc, and the induced potential, $\phi^{ind}$, as given by Eq. (1).

   For  the case of a SWNT TFT device, we are interested  in
a  solution  that corresponds to the case of a uniform
$\phi^{xt}$:  $\phi^{xt}_i=\overline{\phi}, \; \;
\rho_i=\overline{\rho}$. Thus, $\rho$ is the same for each tube in
the array and  can  be written as:
$\overline{\rho}=-C_Q(\overline{\phi}+\sum_n
C_n^{-1}\overline{\rho})$  where $C_n^{-1}$ is the reciprocal
geometric capacitance between  a  single tube and its n-th
neighbor in  the  given array  geometry.  The total induced
potential  is $\phi^{ind}=\overline{\rho} \sum_j C_{ij}^{-1}$ and
the total  reciprocal capacitance of the tube is the sum of  all
$C_n^{-1}$ plus  $C_Q^{-1}$.   The exact analytical expressions
for $C_n^{-1}$ and the sum $C_\infty^{-1}=\sum_n C_n^{-1}$ for
each tube in a regular array of SWNTs separated by  the distance
$\Lambda_0$ can be derived as,
\begin{equation}
C_\infty^{-1}=\frac{1}{\epsilon}\left(2 \log
\frac{2d}{R}+2\sum\limits_{n=1}^\infty
\log\frac{\Lambda_n^2+(2d)^2}{\Lambda_n^2}\right)$$$$=
\left(\frac{2}{\epsilon}\log \frac{\Lambda_0}{R}\frac{\sinh(\pi
2d/\Lambda_0}{\pi} \right)
 \label{eq3}
\end{equation}
where  $\Lambda_n$ is the distance between a given tube and its
n-th neighbor.  To apply this result to the problem of  SWNT  TFT
we  calculate the total charge per unit area induced in  the array
under $\overline{\phi}$:
\begin{equation}
C_i=\frac{Q}{\overline{\phi}S}
=\frac{\overline{\rho}}{\overline{\phi}\Lambda_0}
=\frac{1}{\Lambda_0}\left(C_Q^{-1}+\sum\limits_{n=1}^\infty
C_n^{-1} \right)^{-1}$$$$= \left\{\frac{2}{\epsilon}\log\left[
\frac{\Lambda_0}{R}\frac{\sinh(\pi 2d/\Lambda_0}{\pi}\right]
+C_Q^{-1}\right\}\Lambda_0^{-1}
 \label{eq4}
\end{equation}
$C_\infty$  and  $C_i$  depends on two characteristic lengths:
$x=2\pi d/\Lambda_0$, which  defines  the  inter-tube coupling,
and  $2d/R$,  which defines  the  coupling of a single tube to the
gate.   The physics   of  
 $C_i$ coupling  is
clearly different  for  two  different  regimes  determined   by
$x$ (assuming that $R$ is always the smallest quantity).   In  the
limit   $x \ll  1$ (i.e. sparse tube density) the planar  gate
contribution dominates, $\sinh x \sim x$, and $C_\infty$   reduces
to that for a  single,  isolated tube.  $C_i$ is 
 approximately equal to the product of the capacitance of a
transistor that uses  a  single isolated SWNT with the number of
tubes  per screening  length $\Lambda_0$.  In the opposite limit,
$x \gg 1$, $C_i$ approaches  that of a parallel plate primarily
due  to  the higher surface coverage of tubes.  Meanwhile,
$C_\infty$ decreases due to  the  screening  by neighboring tubes.
To compare  the performance  of SWNT network/array based TFTs with
that  of conventional TFTs that uses continuous, planar channels,
we calculate the ratio of the capacitance of the SWNT-array  to
that  of a parallel plate (Fig. 2). This capacitance  ratio,
$\Xi$, is close to unity for $x\gg 1$:
\begin{equation}
\Xi=1-\frac{\Lambda_0}{2d\pi}\frac{\epsilon}{2}
\left(C^{-1}+C+Q^{-1}-\log 2\right)+\dots \simeq 1
\label{eq5}
\end{equation}
where  $C^{-1}=(2\log (2d/R))/\epsilon$  is  the  single  tube
capacitance. The term in parenthesis is multiplied by the inverse
tube density $1/x=\Lambda_0/2\pi d$  and is, therefore, negligible
when the $\Lambda_0$ becomes much smaller  than  $d$.  For $x\ll
1$, in the opposite  limit,  $\Xi$   is small and grows linearly
with $d$ as given by:
$\Xi=\frac{2d\pi}{\Lambda_0}\frac{2}{\epsilon}\left(C^{-1}+C_Q^{-1}\right)+\dots$.

   To  verify  certain  aspects of  these  calculations,  we
performed    finite-element-method    (FEM)    electrostatic
simulations (FEMLab, Comsol, Inc) to determine the classical $C_i$
of  the  same model system, in which the induced  charge
distributes itself to establish an equal potential over  the
nanotubes.   In  these calculations, $R$ was set  to  0.7  nm,
which  corresponds to the average radius of SWNTs formed  by
chemical   vapor  deposition.\cite{12}   We  chose
$\epsilon_r=4.0$. The
computations,  shown  in  Fig. 2, agree  qualitatively  with
those  determined by Eq. 4, with deviations  that  are  most
significant  at  small  gate  dielectric  thicknesses  where
quantum effects are significant.

   In  experimentally achievable SWNT TFTs,  the  SWNTs  are
not  spaced equally and, except in certain cases,  they  are
 completely  disordered in the form of  random
networks.\cite{5,13}
To  estimate qualitatively the influence of uneven spacings,
we  constructed  an array composed of five hundred  parallel
SWNTs  with a normal distribution of $\Lambda_0 =100\pm 40$ nm
(Fig. 3a inset). $C_i$  was  calculated  by inverting the  matrix
of potential
coefficients  (Eq.  1).   The small difference  of  computed
capacitances  ($\Delta C$)  indicates that  Eq.4  can  be  used
for
aligned arrays with uneven spacings, and perhaps even random
SWNT  networks (Fig. 3).  Another experimental fact is  that
most  SWNT TFTs are constructed in the bottom-gate structure
where  nanotubes  are in an equilibrium distance  above  the
gate dielectric, $\sim 4$\AA,\cite{14} due to van der Waals
interactions.
To  account  for the effect of low $\epsilon$ air medium  on
$C_i$, we performed  the  FEM  simulation for nanotube  arrays
either fully  embedded in gate dielectric or fully exposed  in
the air.   Comparing the capacitances in these two  cases  shows
that the low $\epsilon$ air medium has most significant influence  on
those  systems  that use high $\epsilon$ dielectrics because  of  the
higher dielectric contrast.   Moreover, at $x \ll 1$ the effect
of  the air on SWNT arrays is close to results obtained  for
devices  based  on individual tubes (Fig.3  b).\cite{15}
However,
with  increasing $x$, the screening between neighboring  tubes
forces  electric field lines to terminate on the  bottom  of
nanotubes without fringing through the air and thus the  air
effect diminishes.

   To  explore  these effects experimentally, we built  TFTs
that   used   random  networks  of  SWNT  with  fixed
$1/\Lambda_0$ (approximately  10 tubes/micron, as evaluated  by
AFM)  and
different $d$.  Details on the device fabrication can be found
elsewhere.\cite{16}   For  the range of $\Lambda_0$ and $d$
 here the difference in $C_i$ that results  from  the
air  effect  is less than 20\%, smaller than the experimental
error  in  determining transconductance ($g_m$).  
 So, Eq.4  gives sufficiently accurate estimation of $C_i$.
Figure 4a  shows  the  transfer curves of  SWNT  TFTs.   Figure 4c
compares  effective mobilites calculated  using  $C_i$  derived
from  parallel plate model ($\mu_p$
 ) to
those that from Eq.4 ($\mu$). Consistent  with the previous
discussion, the parallel-plate capacitor model overestimates $C_i$
 significantly for low  tube  densities/thin dielectrics. As  a
result,  the effective mobilities calculated in this manner
($\mu_p$) have  an apparent linear dependence on $d$ that derives
from inaccurate values  for  $C_i$.   On  the other hand,
effective  mobilities calculated using Eq.4 ($\mu$) show no
systematic change with $d$, which provides a validation of the
model.

   The  computed  capacitances  also  reveal  two  important
guidelines  for the development of SWNT TFTs.   First,  Eq.3
and  Eq.4, indicate that the effective $\mu$ should be close  to
the  intrinsic mobility of SWNTs ($\mu_{per tube}$) if the
contact
resistance is neglected since
\begin{equation}
\mu= \frac{L}{W C_i V_{DS}}\left|\frac{\partial I_{DS}}{\partial
V_{GS}}\right|=
\frac{\left(C_\infty^{-1}+C_Q^{-1}\right)L}{V_{DS}}
\left|\frac{\partial I_{DS}/N}{\partial V_{GS}}\right|\simeq
\mu_{per tube} \label{eq6}
\end{equation}
where  $N$  is  the  total number of 
 effective
pathways, connecting source/drain electrodes.  $\mu$  can  be  a
little  smaller than $\mu_{per tube}$ because the actual length  of
effective  pathway  is longer than $L$.  The  huge  difference
between $\mu$ of devices based on SWNT films\cite{5,6,13} and
$\mu_{per tube}$
extracted  from  those  FETs  based  on  individual
tubes\cite{17}
suggests  that  the  tube/tube contacts severely  limit  the
transport  in  SWNT  networks or partially  aligned  arrays,
either  due  to  
 tunneling barrier  
 or 
 electrostatic screening at the contact 
 which  prevents  the  effective  gate  modulation  at   that
specific  point.\cite{18}
Secondly, efforts on
improving
$g_m$  through  increasing  tube density  for  a  given  device
geometry and $V_{DS}$ are limited by $d$.  From Fig. 2a we can
see
that at given $d$, when $x \ll 1$, $C_i$, and thus $g_m$
increases with
the  increase  of tube density.  However,  when  $x  \gg 1$,
$C_i$
saturates  and  $g_m$ no longer increases with  decreasing
$\Lambda_0$.
This prediction was verified by the almost identical $g_m$  for
devices   with  different  tube  densities  on  thick   gate
dielectric  (Fig. 4b). 

   In  summary, we evaluated $C_i$ of SWNT TFTs.
 Our analysis shows that the  best
electrostatic coupling between the gate and the SWNT  occurs
in  dense  arrays  of tubes, but advantage  gained  in
 $C_i$ coupling  from higher tube  density  starts  to
saturate  as $\Lambda_0$ becoming close to $d$.  These conclusions
are corroborated by vertical scaling experiments on SWNT network
TFTs.   We further propose two guidelines for improving  the
performance  of SWNT TFTs.

\section*{ACKNOWLEDGEMENTS}

We  thank T. Banks for help with the processing.  This  work
was  supported by the U. S. Department of Energy under grant
DEFG02-91-ER45439 and the NSF through grant NIRT-0403489.

\newpage
\begin{figure}[htb]
\centering
\includegraphics[width=2.5in]{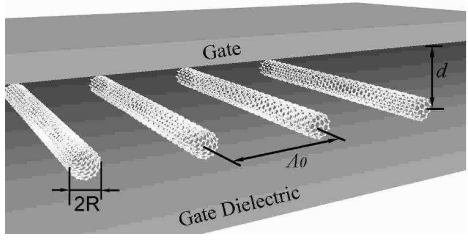}
 \caption{
Schematic illustration of the model system.  $R$ is the nanotube
radius;  the distance between each  tube  and  the dielectric
thickness are  $\Lambda_{o}$ and $d$ respectively. }
\end{figure}

\newpage

\begin{figure}[htb]
\centering
\includegraphics[width=2.5in]{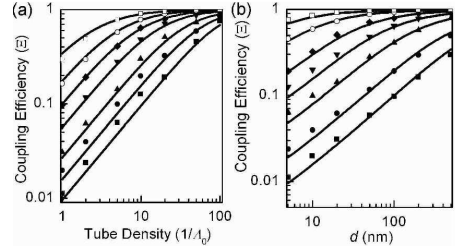}
 \caption{
(a) Capacitance ratio ($\Xi$) calculated by FEM  versus linear
SWNT  densities ($1/\Lambda_{o}$) for $d$ ranging from  5nm
(filled square), 10nm  (filled circle), 20nm (filled triangle),
50nm (upsidedown triangle), 100nm (open square), 200nm (open
circle) to 500nm (small square).  Lines are $\Xi$ calculated
according to Eq.4 for $d$ ranging from 5nm to 500nm, from bottom
to up.  (b) $\Xi$  calculated  by FEM  versus  $d$ for various
$1/\Lambda_{o}$ ranging from 1$\mu$m  (black, filled square),
500nm (filled circle), 200nm (filled triangle), 100nm (upsidedown
triangle), 50nm (filled diamond), 20nm (open circle) to 10nm
(small square).Lines are $\Xi$ calculated according to Eq.4 for
$\Lambda_{o}$ ranging from 1$\mu$m to 10nm, from bottom to up.
 }
\end{figure}

\newpage

\begin{figure}[htb]
\centering
\includegraphics[width=2.5in]{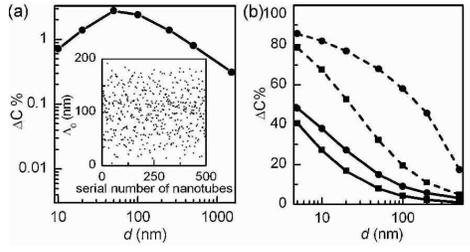}
 \caption{
(a) Relative variation of $C_i$ ($\Delta C$) induced by uneven
 $\Lambda_{o}$  versus $d$.  Inset:  $\Lambda_{o}$ associated with each nanotube in an
array.  (b) Relative difference between $C_i$ of fully embedded
and  fully  exposed nanotube arrays ($\Delta C$)  versus  $d$.
Solid lines  and  dashed  lines represent  results  obtained  from
normal  ($\epsilon_r=4.0$) and high $\epsilon_r$ (15) dielectrics
respectively. Solid circles and squares show results obtained for
$\Lambda_{o}$=100nm and $\Lambda_{o}$=10nm, respectively.
 }
\end{figure}

 ~\newpage
 ~\newpage

\begin{figure}[htb]
\centering
\includegraphics[width=2.5in]{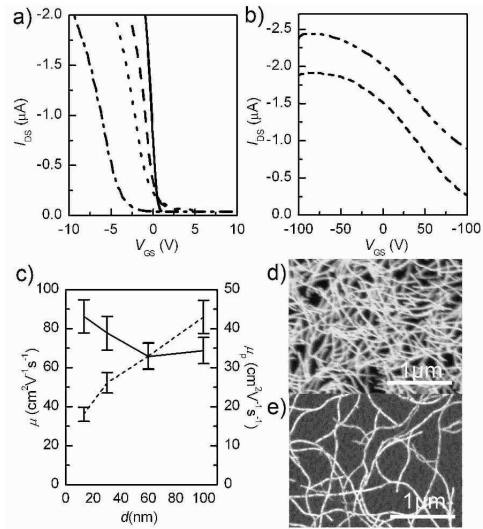}
 \caption{
(Part  a)  Drain/source current  ($I_{DS}$)  versus  gate voltage
($V_{GS}$) at a fixed drain/source bias ($V_{DS}$) of  -0.2  V
collected  from  SWNT TFTs using high density  SWNT  network (SEM
image shown in part d) with a bilayer dielectric of 3nm HfO2 layer
and  an overcoat epoxy layer of  10  nm  (solid line), 27.5
nm(dash line) and 55nm (dot line) thickness,  or a  single  layer
100 nm SiO2 (dash dot line). (Part  b)  $I_{DS}$ versus  $V_{GS}$
collected from SWNT TFTs using high density network  (shot dash
line) and low density SWNT network  (SEM image  shown  in part e)
(dash dot dot line) with  a  single layer 1.6$\mu$m epoxy
dielectrics. Devices have channel lengths ($L$)  of  100 $\mu$m
and effective channel widths ($W$)  of  125$\mu$m. (Part  c) $\mu$
computed based on parallel plate and SWNT  array models  for $C_i$
($\mu_p$,  solid  line,  and  $\mu$,  dashed   line,
respectively).
 }
\end{figure}

\end{document}